\documentclass[prd,superscriptaddress,a4paper,showpacs,showkeys,11pt,nofootinbib]{revtex4}
\usepackage{graphicx}
\usepackage{amsmath}
\usepackage{color}

\begin{document}
\title{Probing Lorentz violating effects on the exclusive $e^+e^-$ production in ultraperipheral heavy - ion collisions}

\author{Laura {\sc Duarte}}
\email{l.duarte@unesp.br}
\affiliation{Institute of Physics and Mathematics, Federal University of Pelotas (UFPel), \\
  Postal Code 354,  96010-900, Pelotas, RS, Brazil}

\author{Victor P. {\sc Gon\c{c}alves}}
\email{barros@ufpel.edu.br}
\affiliation{Institute of Physics and Mathematics, Federal University of Pelotas (UFPel), \\
  Postal Code 354,  96010-900, Pelotas, RS, Brazil}

\author{Daniel E. {\sc Martins}}
\email{daniel.ernani@ifj.edu.pl}
\affiliation{The Henryk Niewodniczanski Institute of Nuclear Physics (IFJ)\\ Polish Academy of Sciences (PAN), 31-342, Krakow, Poland
}

\begin{abstract}
The impact of Lorentz violating (LV) terms on the exclusive $e^+e^-$ production in ultraperipheral heavy-ion collisions at the Large Hadron Collider (LHC) is investigated considering vectorial and axial couplings. Results for the differential and total cross - sections are presented and the sensitivity to a time - like coupling is estimated. Our results indicate that this process can improve the current upper bounds on the LV terms.

\end{abstract}


\maketitle

\section{Introduction}

The Standard Model (SM) of particle physics, theoretically based on Gauge and Lorentz symmetries, has enjoyed remarkable success over the past decades, consistently validated by experimental evidence supporting its theoretical predictions \cite{ParticleDataGroup:2020ssz}. However, it is recognized not as a definitive theory but rather as a low energy effective approximation of a more fundamental high energy theory.

In certain scenarios for the beyond the Standard Model (BSM) physics, Lorentz symmetry can be violated, leading to the emergence of effective low-energy interactions. 
In this context, the so-called Standard Model Extension (SME), conceived by Colladay and Kosteletck\'y \cite{Colladay:1998fq,Colladay:2001wk}, has a comprehensive list of possible Lorentz 
violating (LV) terms, which can be incorporated into all sectors of interactions. 
Since then, the analysis of the effects of Lorentz symmetry breaking has become important in the most diverse areas, from string theory \cite{Kostelecky:1988zi}, supersymmetric models \cite{Berger:2001rm, Carlson:2002zb} and higher-derivative field theories \cite{Gomes:2009ch, Mariz:2011ed} to condensed matter systems \cite{Belich:2006pi, Kostelecky:2021bsb} and beyond. 
In this study, we focus specifically on its implications within the realm of quantum electrodynamics (QED) scattering processes \cite{Colladay:2001wk, Charneski:2012py,deBrito:2016zav, Celeste:2024duc}, where the LV terms imply the modification of the fermion-photon vertex and, consequently, has direct impact on the photon - induced interactions that are being studied in ultraperipheral heavy-ion collisions at the LHC (for a review see Ref. \cite{upc}).

\begin{figure}[t]
{\includegraphics[width=0.55\textwidth]{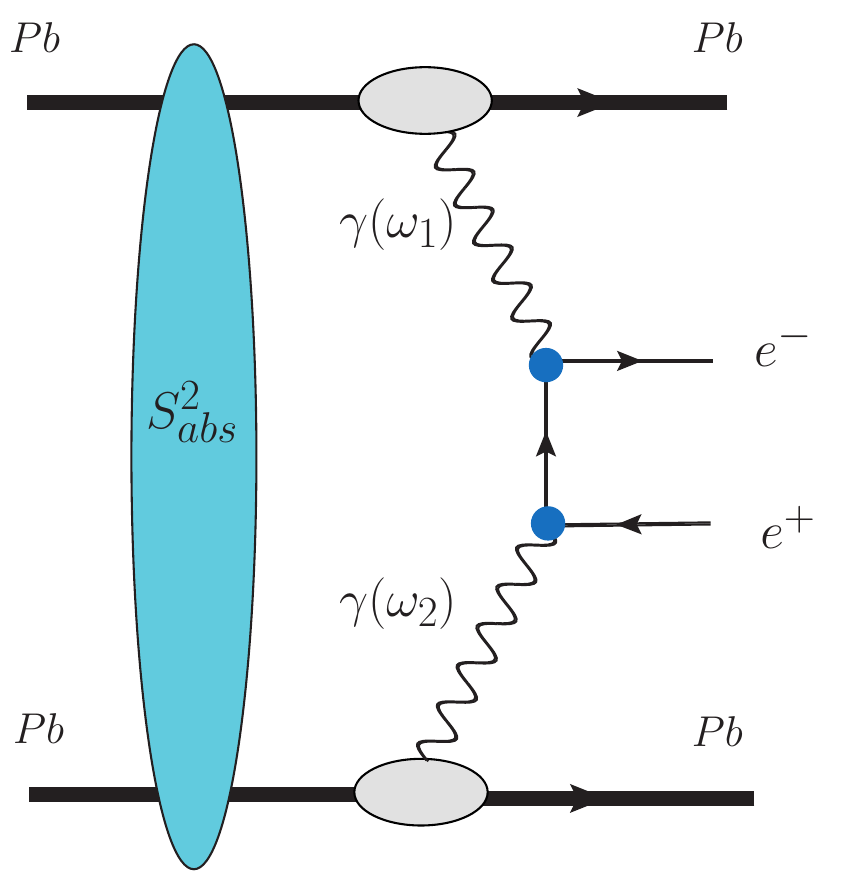}} 
\caption{Dielectron production in ultraperipheral $PbPb$ collisions. The blue blob indicates an electron - photon vertex modified by a Lorentz violating (LV) term. }
\label{Fig:diagram}
\end{figure}

Ultraperipheral heavy-ion collisions (UPHICs) are characterized by an impact parameter $b$ greater than the sum of the radius of
the colliding nuclei,  which implies the suppression of strong interactions and the dominance of the electromagnetic interaction between them \cite{upc}. 
The electromagnetic fields associated with the ultrarelativistic nuclei can be treated as fluxes of quasi-real photons according to the equivalent photon approximation (EPA) formalism \cite{Budnev:1975poe}, with the photon flux being proportional to the square of the nuclear charge $Z$, which implies that the photon - photon ($\gamma\gamma$) cross - sections are enhanced by a factor $Z^4$ in heavy-ion collisions.
 For $PbPb$ collisions,  this $Z^4$ enhancement amounts approximately  to $4.5\times 10^7$, which has allowed us to observe, for the first time, the light-by-light scattering \cite{ATLAS:2019azn,CMS:2018erd}. Here, we will focus on another of the most fundamental QED processes, the pair production in $\gamma \gamma$ interactions, which is characterized by a larger cross - section and which also have been extensively investigated in UPHICs at RHIC and LHC (For reviews and recent studies on the subject see, e.g., Refs. \cite{Baur:1998ay,Baur:2007zz, Klusek-Gawenda:2010vqb, Azevedo:2019fyz,Goncalves:2020btj,
Wang:2021kxm,Shao:2023zge,Shi:2024gex}).  In the EPA approach \cite{upc,Budnev:1975poe} and assuming the impact parameter representation, one has that the process can be represented by the diagram presented in Fig. \ref{Fig:diagram}, and the  total cross-section for the production of a dielectron pair with rapidity $Y$  is  given by
\begin{eqnarray}
\sigma \left(Pb Pb \rightarrow Pb \otimes e^+ e^- \otimes Pb;s_{NN} \right)   
&=& \int \mbox{d}^{2} {\mathbf b_{1}}
\mbox{d}^{2} {\mathbf b_{2}} 
\mbox{d}Y
\mbox{d}W  \frac{W}{2} \,   N\left(\omega_{1},{\mathbf b_{1}}  \right )
 N\left(\omega_{2},{\mathbf b_{2}}  \right ) \nonumber \\ 
 & \times & S^2_{abs}({\mathbf b})  \hat{\sigma}\left(\gamma \gamma \rightarrow e^+ e^- ; 
W \right )
  \,\,\, ,
\label{cross-sec-2}
\end{eqnarray}
where $\sqrt{s_{NN}}$ is center - of - mass energy of the $PbPb$ collision, $\otimes$ characterizes a rapidity gap in the final state and 
$W = \sqrt{4 \omega_1 \omega_2}$ is the invariant mass of the $\gamma \gamma$ system, which is equal to the invariant mass of the $e^+e^-$ pair ($m_{e^+e^-}$).  Moreover, $N(\omega_i, {\mathbf b}_i)$ is the equivalent photon spectrum
of photons with energy $\omega_i$ at a transverse distance ${\mathbf b}_i$  from the center of the nucleus, defined in the plane transverse to the trajectory. The spectrum can be
expressed in terms of the nuclear form factor \cite{Azevedo:2019fyz}.  
 The factor $S^2_{abs}({\mathbf b})$ depends on the impact parameter ${\mathbf b}$ of the $PbPb$ collision and  is denoted the absorptive  factor, which excludes the overlap between the colliding nuclei and allows taking into account only ultraperipheral collisions.
 The photon energies $\omega_1$ and $\omega_2$  are related to   
$W$ and to the rapidity  $Y = \frac{1}{2}(y_{e^+} + y_{e^-})$ of the outgoing dielectron pair system by 
$ \omega_1 = \frac{W}{2} e^Y$  and $\omega_2 = \frac{W}{2} e^{-Y}$. 
Finally, the cross-section $\hat{\sigma}$ is the elementary cross-section to produce an $e^+ e^-$ pair of leptons with mass $m_e$ in a two - photon interaction, which can be calculated in the SM using the Breit - Wheller formula (For details see, e.g., Ref. \cite{Azevedo:2019fyz}). On the other hand, in the SME approach \cite{Colladay:1998fq,Colladay:2001wk}, such cross - section will be modified by Lorentz violating interactions, which change e.g. the electron - photon vertex, and the associated predictions will be sensitive to the magnitude of the LV coefficients. Our goal in this paper is to investigate possible effects of a LV term on dielectron production in UPHICs and derive the upper bounds on the breaking parameters considering the recent LHC results \cite{ALICE:2013wjo,ATLAS:2022srr}, which have demonstrated that this process is quite well described by the Standard Model.

This paper is organized as follows. In the next section, we will present a brief review of formalism used to introduce LV interactions in the {$e^{+}e^{-}$ } production. In particular, we will consider the possibility of vectorial and axial nonminimal couplings. In Section \ref{sec:res} we will estimate the impact of these LV terms on the total cross-section and associated distributions for the dielectron production in ultraperipheral $PbPb$ collisions at $\sqrt{s} = 5.5$ TeV. Moreover, we will estimate the corresponding sensitivity and derive upper bounds on the LV parameters considering different cuts on the dielectron invariant mass and distinct magnitudes for the systematic uncertainty. Finally, in Section \ref{sec:sum} we will summarize our main results and conclusions.


\section{Lorentz violation with nonminimal couplings}

The Standard Model Extension (SME) contains the standard model and general relativity, as well as all possible Lorentz violating terms that can be constructed from the associated fields (For a review see, e.g., Ref. \cite{Tasson:2014dfa}). These breaking terms can be implemented on the kinetic sector, which implies e.g. the modification of the energy - momentum relation, or in the interaction part via a nonminimal coupling. As in this paper we are interested in the investigation of the LV effects on the $\gamma \gamma \rightarrow e^+ e^-$ process, we will consider the nonminimal coupling approach. In this case, the  Lorentz violation  is introduced by modifying the electron-photon vertex, through the inclusion of a nondynamical background 4-vector $b^{\mu}$, which couples nonminimally to the electromagnetic field strength $F_{\mu\nu}$ (or its dual, $\tilde{F}_{\mu\nu} = \frac{1}{2}\varepsilon_{\mu\nu\alpha\beta}F^{\alpha\beta}$), as well as to the electromagnetic current. 
It is important to emphasize that Lorentz symmetry dictates that physical laws remain unchanged under rotations and boosts, but the presence of $b_{\mu}$
  suggests a violation of this symmetry by privileging a specific direction in space - time.

The inclusion of a nonminimal coupling implies that ${\cal{L}}_{QED} \rightarrow {\cal{L}}_{QED} + {\cal{L}}_{LV} $. In our analysis, we will consider that the nonminimal coupling can be vectorial or axial, and that the associated lagrangians are given by
\begin{equation}\label{vectorial}
  \mathcal{L}_{LV}^{\text{Vectorial}} =   gb^{\nu} \bar{\psi} \gamma^{\mu}\psi \tilde{F}_{\mu\nu} \,\,,
  \end{equation}
and 
\begin{equation}\label{axial}
  \mathcal{L}_{LV}^{\text{Axial}} =  g b^{\nu} \bar{\psi} \gamma^{\mu}\gamma^5 \psi \tilde{F}_{\mu\nu} \,\,,
  \end{equation}
where $g$ is the effective coupling constant, $b^{\nu} = (b_0, \vec{b})$ is the 4-vector that breaks the Lorentz symmetry, with the term $gb^{\nu}$ having canonical dimension of inverse of mass.
Since $b^{\mu}$ is fixed, it can be characterized as a nondynamical background and Lorentz symmetry is broken, as it selects a privileged direction in space-time. The associated Feynman rules can be obtained from the new lagrangian, and the scattering amplitude at the tree level for the traditional QED processes  can  be directly derived. The comparison of the corresponding cross-sections with the current experimental data allow deriving upper bounds on the magnitude of $gb^{\nu}$. It is useful in the literature to assume that $b^{\nu}$ is timelike ($\vec{b} = \vec{0}$) or  spacelike ($b_0 = 0$). In our analysis, we will focus on the timelike case, what allow us to directly compare  our predictions with the previous results in the literature.
In recent years, the impact of LV terms in QED processes was estimated by several theoretical groups and some important results were obtained: in Ref. \cite{deBrito:2016zav} the authors calculated the effects of nonminimal vectorial coupling on the cross-section related with Compton
and Bhabha scatterings, as well as electron-positron annihilation. They found an upper bound of $gb_0 \lesssim 10^{-3}$ GeV$^{-1}$ related to Bhabha cross-section.
For the $e^+ e^- \to \gamma \gamma$ process, the limit for a time-like background was obtained in  Ref. \cite{Celeste:2024duc} and corresponds to $gb_0 \lesssim 10^{-6}$ GeV$^{-1}$.
The axial coupling was analyzed in Ref. \cite{Charneski:2012py} and 
 an upper bound to the breaking parameter corresponds to $g b_0 = 10^{-3}$ GeV$^{-1}$ for Bhabha scattering\footnote{The authors of Ref. \cite{Charneski:2012py} also analyzed the upper limit for a space-like background and found that $g\vec{b} \lesssim 10^{-5}$ GeV$^{-1}$.}. 
The bounds mentioned above were obtained by comparing the theoretical results of the differential cross-section for QED and for QED plus the LV term with the results of the experimental study performed in Ref. \cite{Derrick:1986de} and assuming an educated guess for a QED-cutoff parameter, which parametrizes possible deviations from the theoretical predictions. As the correct value for this quantity is still an open question,  the derived bounds are impacted by the associated uncertainty.

In this paper, we will  extend the approach used in these previous studies for the $\gamma \gamma \rightarrow e^+ e^-$ process. One has that the scattering amplitude can be written in a schematic way, as follows
\begin{eqnarray}
    i{\cal{M}}_{total} = i{\cal{M}}_{0} + i{\cal{M}}_{1} + i{\cal{M}}_{2} \,\,,
\end{eqnarray}
where the index indicates the power of $g b^{\nu}$ present in the amplitude, i.e. $i{\cal{M}}_{0}$ does not depend on $g b^{\nu}$ and is the usual QED prediction. On the other hand, $i{\cal{M}}_{1}$ is linear 
in $g b^{\nu}$, being associated with a diagram where a usual vertex and another with the LV term are present. Finally, $i{\cal{M}}_{2}$ corresponds to the diagram where two LV vertex are present, and is proportional to the second power of $g b^{\nu}$. We will use ${\cal{M}}_{total}$ to estimate the corresponding cross - section ${\sigma}(\gamma \gamma \rightarrow e^+ e^-)$, which is one of the main ingredients to calculate the exclusive dielectron production  in UPHICs [See Eq. (\ref{cross-sec-2})].

\begin{figure}[!t]
\includegraphics[width=14cm]{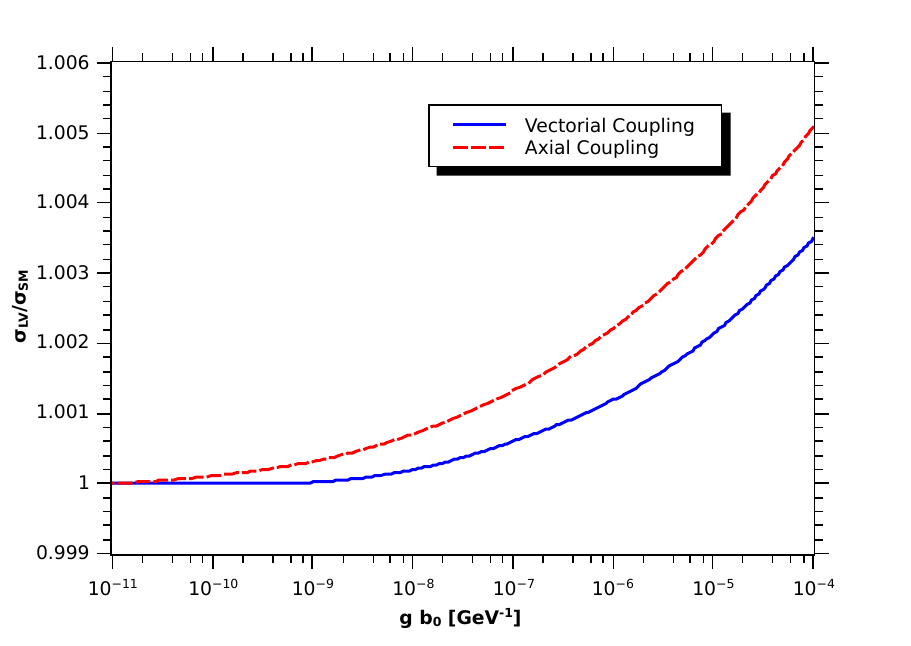}
\caption{Dependence on the time - like coupling $g b_0$ of the ratio between the total cross-section including the LV terms ($\sigma_{LV}$), for the vectorial and axial couplings, and that associated with SM ($\sigma_{SM}$). }
\label{fig:total}
\end{figure}

\section{Results}
\label{sec:res}

In what follows, we will present our results for the exclusive $e^+ e^-$ production in ultraperipheral $Pb\, Pb$ collisions at $\sqrt{s} = 5.5$ TeV. 
The LV models were implemented in the \verb+Feynrules+ package \cite{Alloul:2013bka} to generate a Universal FeynRules
Output (UFO) module \cite{Degrande:2011ua}, which was used in the \verb+MadGraph5+ \cite{Alwall:2011uj} to generate the events. In our analysis, we assume the point - like nuclear form factor to estimate the corresponding photon fluxes and that the absorptive factor $S^2_{abs}(b)$ in Eq. (\ref{cross-sec-2})  is equal to zero for $b < 2R$ and equal to 1 for $b \ge 2R$, where $R$ is the nuclear radius. A detailed discussion about the dependence of the predictions on these assumptions is presented in Ref.  \cite{Azevedo:2019fyz}, which we refer to the interested reader. In Fig. \ref{fig:total} we present our results for the dependence on $g b_0$ of the ratio between the total cross-section including the LV terms ($\sigma_{LV}$) for the vectorial and axial couplings and that associated with the SM ($\sigma_{SM}$). As expected, the ratio becomes equal to 1 for very small values of $g b_0$. On the other hand, the ratio increases for large values of $g b_0$, with the impact being larger in the axial case. Another important aspect is that for $g b_0 \ge 10^{-8}$, the ratio is larger than $1.0005$. As we will demonstrate in what follows, this small deviation from the unity already allow us to derive strong upper bounds on the values of the couplings, since the cross - sections for the exclusive $e^+ e^-$ production in UPHICs at LHC are of the order of mb and the integrated luminosity per year is $\mathcal{L} = 1.72$ nb$^{-1}$.

\begin{figure}[t]
	\centering
	\begin{tabular}{cc}
    \includegraphics[width=0.5\textwidth]{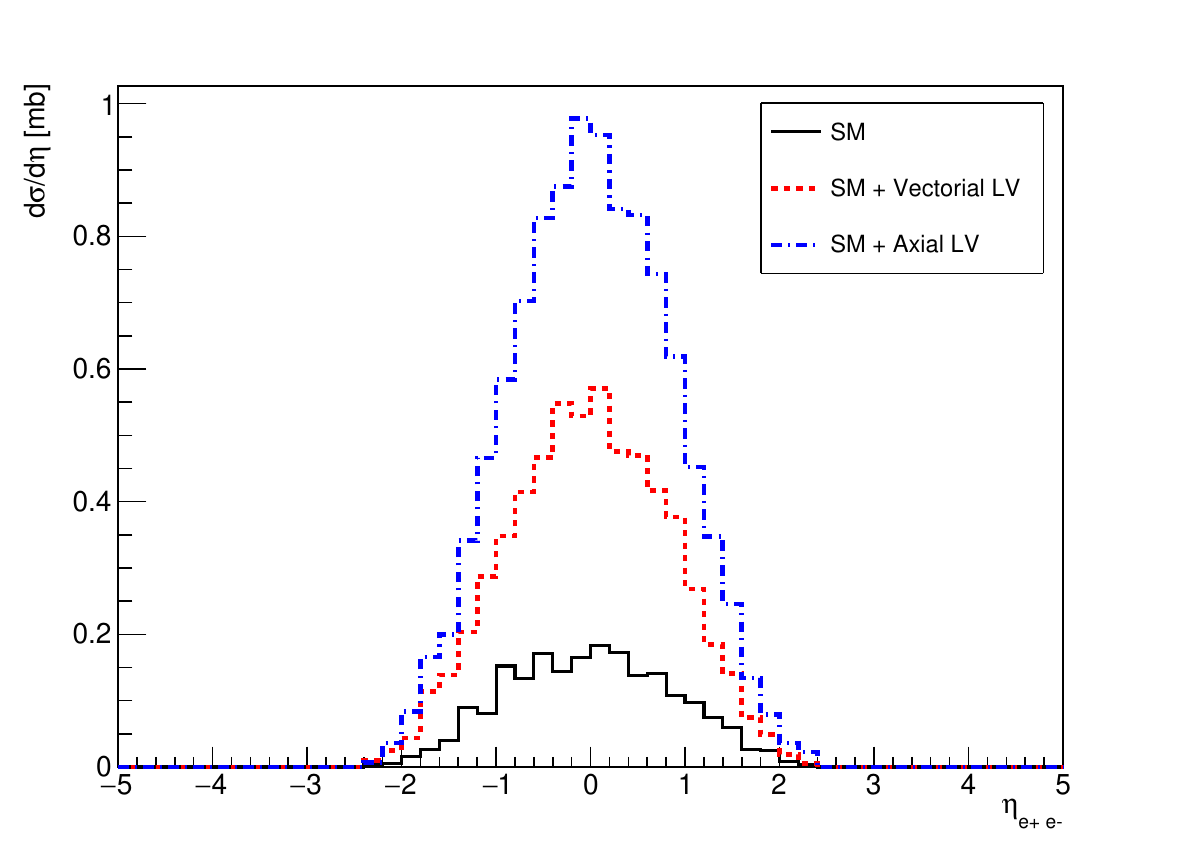} &
    \includegraphics[width=0.5\textwidth]{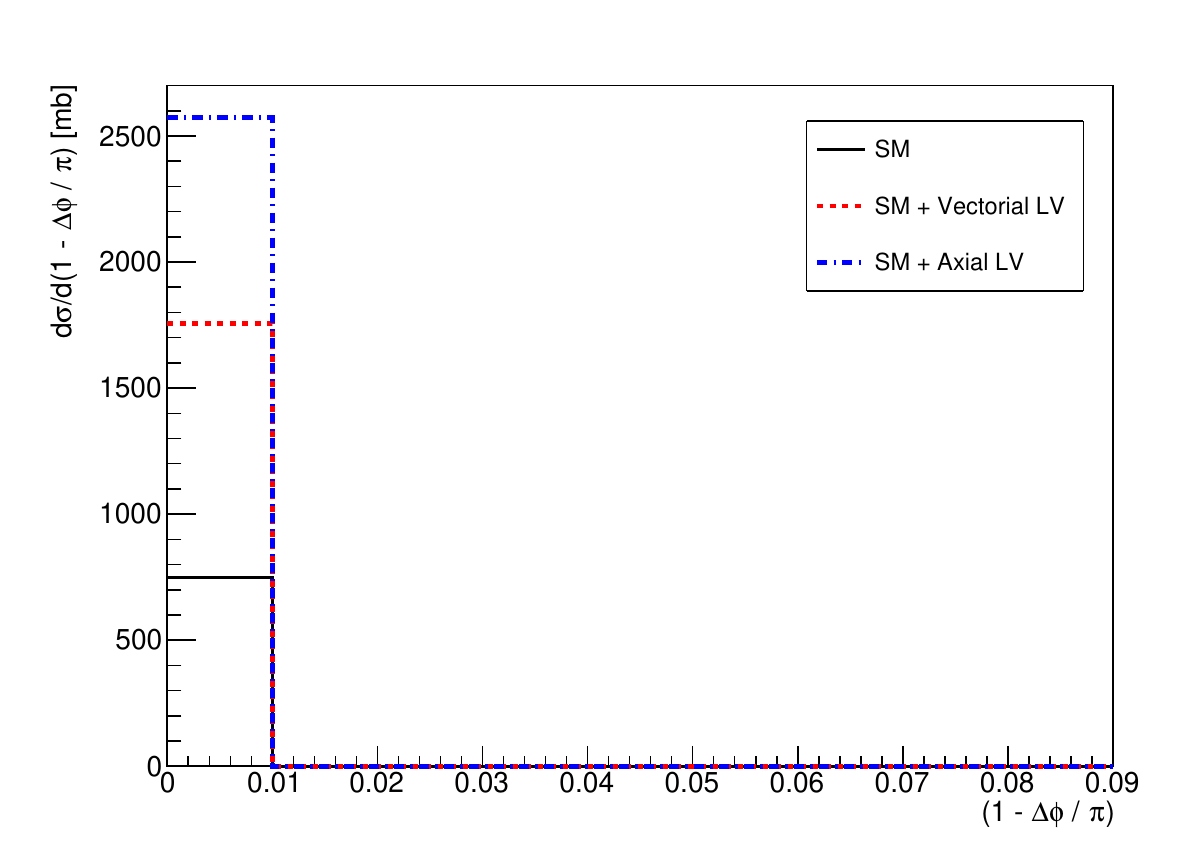} \\\includegraphics[width=0.5\textwidth]{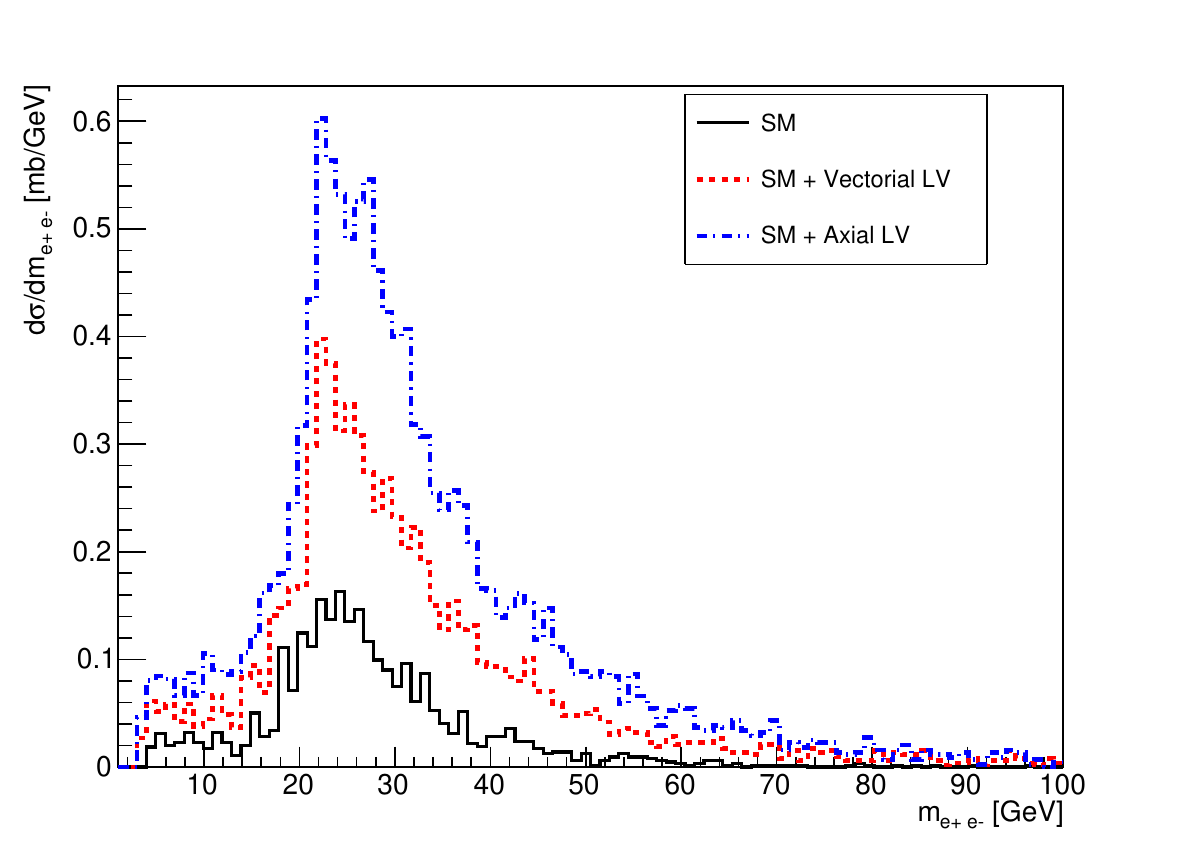} &
    \includegraphics[width=0.5\textwidth]{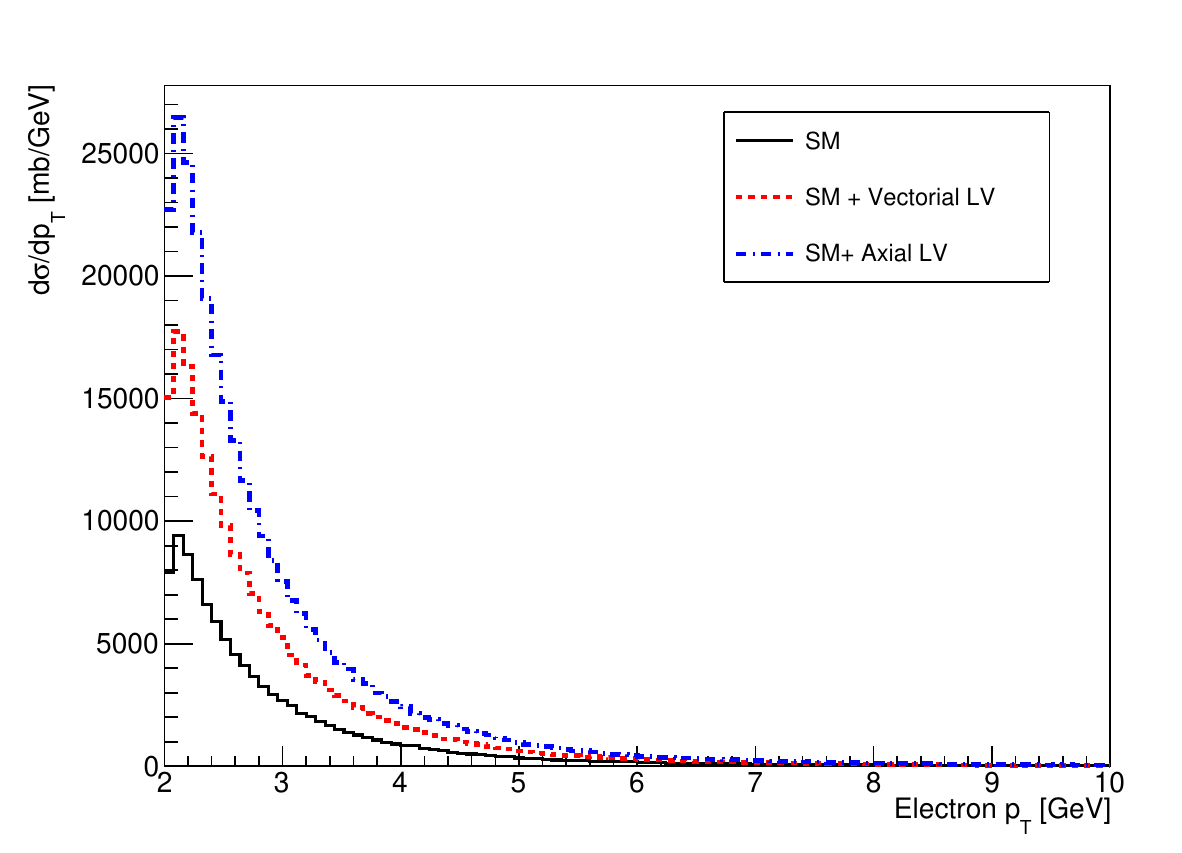}     
			\end{tabular}
\caption{Results for the pseudorapidity ($\eta_{e^+ e^-}$), acoplanarity ($1 - \Delta \phi/\pi$), invariant mass ($m_{e^+ e^-}$) and transverse momentum ($p_T(e)$) distributions, derived assuming  $g b_0 = 10^{-5}$ and $\sqrt{s_{NN}} = 5.5$ TeV}
\label{fig:distribuicoes}
\end{figure}

Figure \ref{fig:distribuicoes} presents our predictions for the pseudorapidity ($\eta_{e^+ e^-}$), acoplanarity ($1 - \Delta \phi/\pi$), invariant mass ($m_{e^+ e^-}$) and transverse momentum ($p_T(e)$) distributions, derived assuming  $g b_0 = 10^{-5}$. One has that the pseudorapidity and acoplanarity distributions associated with the LV results are similar to those predicted by the SM, differing only in the normalization. On the other hand, the invariant mass and transverse momentum distributions become wider when the LV terms are included. These results motivate us to consider kinematical cuts in these variables. 

Table \ref{tab:1} presents the associated predictions for the total cross-sections corresponding to the LV and SM cases considered in our analysis. Moreover, the impact of distinct kinematical cuts on the predictions is also shown. We assume the pseudorapidity range covered by a central detector ($|\eta_{e^+ e^-}| <2.4$) and the typical cut on the electron transverse momentum ($p_T(e) > 4.0$ GeV) used by the experimental collaborations. 
The kinematical cut in $p_T(e)$ has a large impact on the LV and SM results, reducing the predictions in two orders of magnitude. In contrast, the cut on $\eta_{e^+ e^-}$ has a small impact on the predictions, which is expected from the analysis of the results for $d\sigma/d\eta_{e^+ e^-}$ presented in Fig. \ref{fig:distribuicoes}. Finally, as the difference between the LV and SM predictions increases for larger invariant masses, we present our predictions for two different cuts on the invariant mass of the $e^+ e^-$ pair. As we will demonstrate below, the sensitivity to the LV effects depends on the range considered.


\begin{table}[t]
\centering
\scalebox{0.95}{
\begin{tabular}{ |p{4cm}||p{2cm}|p{3.5cm}|p{3cm}|  }
 \hline
 \multicolumn{4}{|c|}{$\mathbf{Pb\, Pb @ 5.5\, TeV} $}  \\ \hline \hline
 
 \multicolumn{1}{|c||}{$Pb\, Pb \rightarrow  Pb\, e^+ e^-\, Pb $} & \multicolumn{3}{|c|}{$\boldsymbol{\sigma}$ \bf {[mb]} for $\mathbf{b_0 g = 10^{-5}}$ [GeV$^{-1}]$}\\
 \hline
 - & {\bf SM} & {\bf Vectorial LV} & {\bf Axial LV} \\
 \hline
 Total cross section [mb] & $0. 58457$ & $ 0.585814$ &$0.586586$ \\ \hline
 $p_T(e^+,e^-) > 4$ GeV & $0.00155262$      &$ 0.00165508$  &$  0.00168773$  \\ \hline
 $|\eta_{e^+ e^-}| <2.4$ & $0.00153391$     & $ 0.00163637$   & $ 0.00167148$ \\ \hline
 $  20 < m_{e^+ e^-} < 50 $ GeV & $ 0.0014357$ & $ 0.00149762$ & $   0.00154988$  \\ \hline
 $  50 < m_{e^+ e^-} < 100 $ GeV & $9.47\times 10^{-5}$  &$1.19506\times 10^{-4}$ &$1.36172\times 10^{-4}$  \\ \hline \hline 
\end{tabular}}
\caption{Predictions for the total cross-sections for the LV and SM process.  The results were estimated considering 
kinematical cuts on the transverse momentum of the electron and on the pseudorapidity and invariant mass of the $e^+ e^-$ pair system.}
\label{tab:1}
\end{table}

\begin{figure}[t]
	\centering
	\begin{tabular}{cc}
    \includegraphics[width=0.5\textwidth]{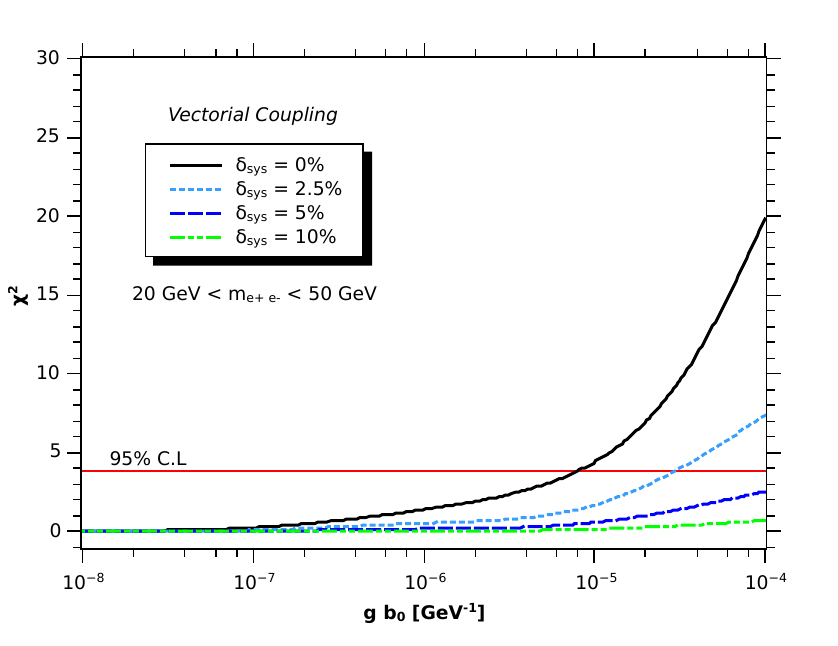} &
    \includegraphics[width=0.5\textwidth]{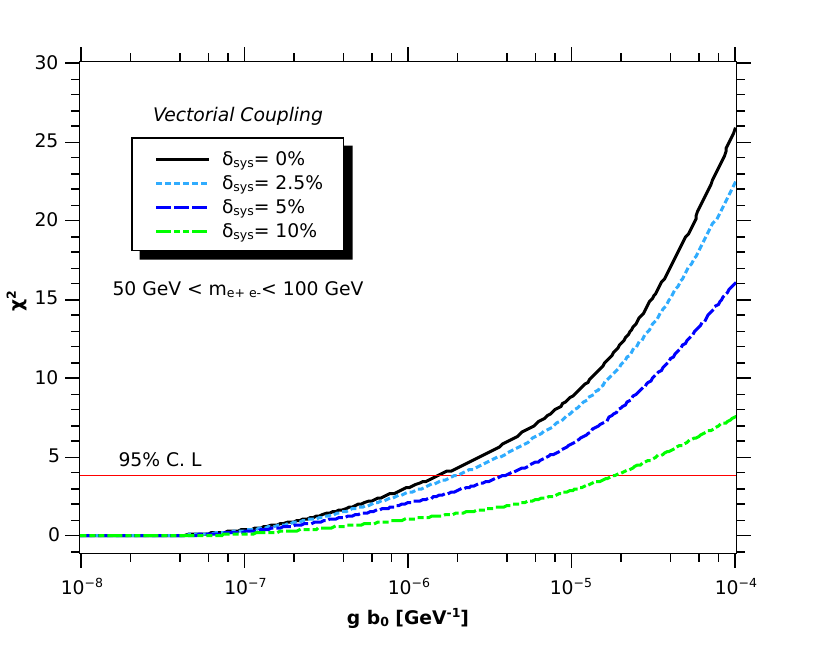} \\
    \includegraphics[width=0.5\textwidth]{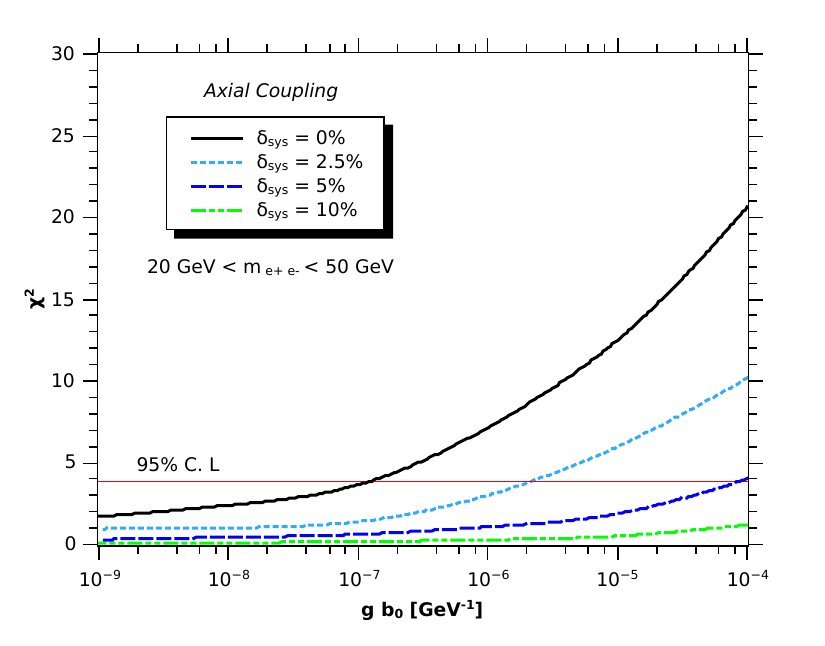} &
    \includegraphics[width=0.5\textwidth]{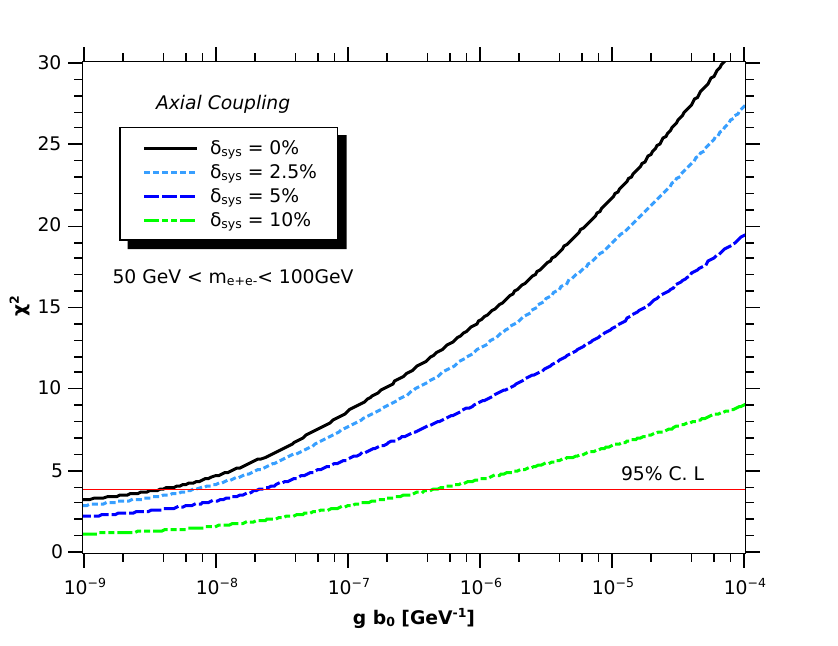} 			
       \end{tabular}
\caption{Sensitivity of the exclusive dielectron production in UPHICs to a Lorentz violating time - like background $b_0 g$. Results for vectorial (upper panels) and axial (lower panels) couplings, two different range of the dielectron invariant mass and distinct amounts for the systematic error. For comparison, the 95 \% confidence level (C.L.) is also presented. }
\label{fig:dist}
\end{figure}

In order to investigate sensitivity of the exclusive dielectron production in UPHICs to a time-like background $b_0 g$, we will use the $\chi^2$ method, with the $\chi^2$  function being defined  as  follows
\begin{eqnarray}
    \chi^2 = \frac{(N_{LV}-N_{SM})^2}{N_{LV}^2\left(\frac{1}{N_{LV}}+\delta_{\text{sys}}^2\right)} \,\,,
\end{eqnarray}
where $N_{SM}$ and $N_{LV}$ are the total number of events associated with SM process and including the LV terms, obtained after the implementation of the cuts shown in Table \ref{tab:1} and using an integrated luminosity of $\mathcal{L} = 1.72$ nb$^{-1}$. Moreover, we will also consider the impact on our results of different possible values for the systematic error $\delta_{\text{sys}}$. In Fig.~\ref{fig:dist} we present our predictions for the  $\chi^2$ as a function of $b_0 g$, derived considering the vectorial and axial couplings and two distinct ranges for the invariant mass of the dielectron pair. As expected from  Fig.~\ref{fig:total}, the process is more sensitive to an axial coupling, independently of  selection in $m_{e^+ e^-}$.  However, for vectorial and axial couplings, the sensitivity is larger considering events characterized by $50\,\mbox{GeV} \le m_{e^+ e^-} \le 100\,\mbox{GeV}$. For this invariant mass range, our results for the vectorial coupling indicate that the exclusive dielectron production is sensitive to values of $b_0 g$ larger than $10^{-6}$ GeV$^{-1}$, with the minimum value being dependent on $\delta_{\text{sys}}$. In contrast, for an axial coupling, the process becomes sensitive to $b_0 g \ge 10^{-8}$ GeV$^{-1}$. These results indicate that the analysis of the exclusive production of $e^+ e^-$ pairs with large invariant mass in UPHICs can improve in two orders of magnitude the upper bounds in the time-like LV background, derived in previous studies that considered QED processes to constrain the LV effects. Although this kinematical range has not yet been investigated in detail by the ATLAS and ALICE collaborations, measurements of dileptons with large invariant masses are expected in the forthcoming years~\cite{dEnterria:2022sut}. On the other hand, the existing data in the range $20\,\mbox{GeV} \le m_{e^+ e^-} \le 50\,\mbox{GeV}$, which can be quite well described by SM predictions, already allow us to exclude  values of $b_0 g$ of the order of $10^{-7}$ GeV$^{-1}$, improving the current upper bounds.   

\section{Summary}
\label{sec:sum}

In recent years, the study of photon - induced interactions in heavy - ion collisions at the LHC became a reality, which has allowed us to probe rare QED processes and improve our understanding of the standard model. In addition, UPHICs also have provided important new constraints in several scenarios of BSM physics. In this paper, we have considered the SME framework, which is characterized by the presence of Lorentz violating terms, and investigated the impact of the LV effects on the exclusive $e^+ e^-$ production in $PbPb$ collisions at the LHC. In particular, we have considered a time - like background and derived the associated differential and total cross - sections for  vectorial and axial couplings. The predictions were compared with the corresponding SM results and the sensitivity to the LV effects was estimated. Our results demonstrated that the current data already allow us to improve the current upper bounds in the couplings, and  a future experimental analysis of the events with large invariant masses will expand the reach for LV effects in two orders of magnitude. The results presented here strongly motivate the investigation of the LV effects on the exclusive dilepton production in proton - proton collisions, which has been measured by the LHC collaborations considering the tagging of the protons in the final states using forward detectors \cite{CMS:2011vma,CMS:2018uvs,ATLAS:2017sfe,ATLAS:2020mve,CMS:2024qjo}, and that probes dileptons with $m_{l^+ l^-} \ge 200$ GeV. The results of this study will be presented in a forthcoming paper.

\begin{acknowledgements}
VPG acknowledges a useful discussion with A. M. Santos, which has motivated the present study.  Daniel Ernani acknowledges the support of POLONEZ BIS project No. 2021/43/P/ST2/02279 co-funded by the National Science Centre and the European Union's Horizon 2020 research and innovation programme under the Marie Skłodowska-Curie Actions grant agreement no. 945339. This work was  partially financed by the Brazilian funding agencies CNPq,   FAPERGS and INCT-FNA (processes number 
464898/2014-5).
\end{acknowledgements}

\end{document}